\documentclass[oneside,a4paper]{article}

\usepackage{hyperref}
\usepackage{amsmath}
\usepackage{amssymb}
\usepackage{pifont}
\usepackage{pgfplots}
\usepgfplotslibrary{fillbetween}
\usepackage[margin=0.25in]{geometry}
\pgfplotsset{width=10cm,compat=1.9}

\usepackage{authblk}
\usepackage{paralist}
\providecommand{\keywords}[1]{\textbf{\textit{Keywords:}} #1}

\newcommand{\cmark}{\ding{51}}%
\newcommand{\xmark}{\ding{55}}%

\newcounter{example}
\newenvironment{example}[1][]{\refstepcounter{example}\par\medskip
  \noindent \textbf{Example~\theexample. #1} \rmfamily}{\medskip}

\title{Getting more from the skyline operator through restricted skylines, regret minimizing sets and skyline ordering: a survey on their properties and comparison}
\author{Claudio Migliorelli}
\affil{Politecnico di Milano\\
  Milan, Italy\\
  \href{mailto:claudio.migliorelli@mail.polimi.it}{claudio.migliorelli@mail.polimi.it} }
\date{}

\begin{document}
\maketitle
\begin{abstract}
  Given a set of multidimensional points, the skyline operator returns a set of potentially interesting points from such a dataset. This popular operator filters out a set of tuples that are not dominated by other ones, reducing the size of a possibly large initial dataset. However, with the dataset dimensions growing, the query result of the skyline operator can have a high cardinality that could overwhelm the final user looking at the result, yielding the very opposite goal. Moreover, it doesn't allow any user to express their preferences over the attributes, in order to control the dataset cardinality and to return a more satisfying result. In this survey, we discuss and compare the more recent approaches developed to extend the skyline operator's functionalities, and we discuss their applicability in real-world scenarios. We show how these frameworks operate on the dataset and how they claim to mitigate the drawbacks coming from the skyline operator.
\end{abstract}


\section{Introduction}
In modern data-intensive scenarios, the availability of representative data points and the quick overview of possibly interesting objects is a crucial task. The problem of optimizing different attributes assigned to the tuples contained in a given dataset goes under the name of \textit{multi-objective optimization}, and it is widely studied \cite{freitas_critical_2004}. Generally, three approaches are adopted:
\begin{inparaenum}[\itshape a\upshape)]
\item the \textit{lexicographical} approach, for which a strict priority between attributes is defined,
\item the \textit{ranking queries} (top-$k$) approach, for which the query result-set is made of all the tuples with the best score according to a scoring function defined over the attributes, and
\item the \textit{skyline} approach, which consists of all the non-dominated tuples in the initial dataset.
\end{inparaenum}\\
The last two approaches are the most used in data-intensive scenarios, even if they have strengths and drawbacks, as stated in \cite{freitas_critical_2004}.\\
In particular, the skyline operator \cite{borzsony_skyline_2001} returns the set of tuples that are not dominated by other tuples: a tuple $t$ dominates another tuple $s$ if $t$ is no worse than $s$ on all attributes, and it is strictly better in at least one of them.\\
Top-$k$ queries \cite{ilyas_survey_2008}, instead, exploit the definition of a so-called \textit{scoring function} that defines the score of a tuple $t$. Choosing a scoring function allows the user to give different importance to different attributes, through the selection of weights parameters: a top-$k$ query returns the best $k$ objects according to such a scoring function.\\
Top-$k$ queries transfer user preferences in a particular choice of weights in the scoring function, but their outcome may vary even with a small difference in the user's parameters. Skyline queries, instead, may offer a broader overview of the possibly interesting tuples, but the outcome cardinality may get bigger than expected.\\
To improve the skyline operator's effectiveness, in \cite{ciaccia_reconciling_2017, ciaccia_flexible_2020}, the authors introduce a new framework, called \textit{restricted}/\textit{flexible skyline} (F-skyline and R-skyline), which reconcile skyline queries with top-$k$ queries. Specifically, using constraints on the weights of linear scoring functions, the framework induces the skyline to return only those tuples belonging to it satisfying the constraints previously defined. To do so, a new ``dominance'' concept, called $\mathcal{F}$-dominance, is introduced: given a family $\mathcal{F}$ of linear scoring functions, a tuple $t$ $\mathcal{F}$-dominates another tuple $s$ when $t$ is no worse than $s$ according to all the scoring functions in $\mathcal{F}$. This approach allows exploiting the strength of both skyline and ranking queries, obtaining a succinct overview of the dataset. In \cite{mouratidis_exact_2018}, Mouratidis et al. give an extension of the R-skyline operators to return the set of tuples that are $\mathcal{F}$ dominated by less than $k$ tuples, along with the set of tuples that appear in the top-$k$ result for some scoring function belonging to the family $\mathcal{F}$. A similar approach can be found also in \cite{mouratidis_marrying_2021}, in which authors introduce the $\operatorname{ORD}$ and $\operatorname{ORU}$ operators. Another natural extension of the flexible skyline concept is the $k$-skyband operator \cite{papadias_progressive_2005}, which returns those tuples that are dominated by at most $k$ other tuples.\\
The approach introduced in \cite{nanongkai_regret-minimizing_2010}, instead, aims to provide a set of interesting points through the \textit{regret minimization} concept. The final goal is to minimize the \textit{maximum regret ratio}, which is related to how disappointed a certain user could be seeing, for instance, $k$ representative tuples instead of the whole dataset. Such a set of $k$ representative tuples minimizing the maximum regret ratio takes the name of $k$-regret minimizing set.
In \cite{nanongkai_regret-minimizing_2010}, Nanongkai at al. introduce the $k$-regret operator and its properties, hybridizing the skyline operator with top-$k$ queries, and point out that, on the theoretical side, computing an optimal solution may be NP-hard. In \cite{chester_computing_2014}, Chester et al. introduce a relaxation of the $k$-regret operator, showing that, even with such a relaxation, the problem is NP-hard. In \cite{cao_k-regret_2017}, Cao et al. propose efficient algorithms to compute $k$-regret minimizing sets in $2$ dimensions and approximation algorithms in $3$ or more dimensions. In \cite{wang_fully_2021}, Wang et al. present a fully dynamic algorithm to transpose the solutions studied before, which are all related to static settings, in a dynamic setting to maintain the result efficiently even when the database is updated. In \cite{xie_experimental_2020}, Xie et al. analyze the latest state-of-the-art regret minimization techniques and they provide an experimental comparison between them on both synthetic and real datasets.\\
To introduce a constraint on the skyline cardinality, in \cite{lu_flexible_2011}, Lu et al. introduce the \textit{size constraints skyline queries}, which take an input $\tilde{k}$ and return $\tilde{k}$ interesting points from a given dataset. These $\tilde{k}$ desired tuples are obtained exploiting the \textit{skyline ordering} approach, which is a skyline-based partitioning of the initial dataset. A similar approach can be found in \cite{vlachou_ranking_2010}. Vlachou et al. define a framework called SKYRANK that arranges skyline points in a graph relying on the dominance relationships. In such a graph, skyline points are ranked according to the well-known PageRank algorithm \cite{brin_anatomy_1998}. A similar approach relying on subspace dominance relationships has been investigated in \cite{chan_high_2006}, whose main idea is to find a fixed-length set of tuples that belong to the most subspace skylines possible.\\
In section \ref{sec:restricted_skylines}, we describe in more detail the restricted skylines concept, highlighting its strengths and applicability. In section \ref{sec:regret_minimizing}, we dive into the details of the regret minimization problem, describing its underlying ideas and extension, while, in section \ref{sec:skyline_ordering}, we describe the solutions given for the skyline ordering technique, and we analyze their effectiveness in real-world scenarios. Finally, in section \ref{sec:conclusions}, we provide a comparison between these three techniques, highlighting their pros and cons.

\section{Restricted/flexible skylines}\label{sec:restricted_skylines}
\subsection{Introduction to restricted/flexible skylines and first definitions}
The main known problem of skyline queries \cite{borzsony_skyline_2001} regards the unconstrained cardinality of the resulting skyline, mainly due to the lack of user preferences. The underlying idea behind \cite{ciaccia_reconciling_2017, ciaccia_flexible_2020} relies on hybridizing the skyline operator with the top-$k$ approach. The goal is empowering skyline queries capturing user preferences using constraints on the weights appearing in the scoring function.\\
Consider a relation schema $R$ with attributes $A_1, \dots, A_d$. A tuple $t$ over $R$ is a function that associates a value $v_i \in [0,1]$ to each attribute $A_i$. The tuple can be denoted as $t=\langle v_1, \dots, v_d \rangle$ and each value $v_i$ can be denoted equivalently as $t[A_i]$.\\
A \textit{scoring function} is a function $f:[0,1]^d \rightarrow \mathbb{R}$. For each tuple $t = \langle v_1, \dots, v_d \rangle$, the value $f(v_1, \dots, v_d)$ is called the \textit{score} of $t$, and can be denoted as $f(t)$.
As defined in \cite{ciaccia_flexible_2020}, a scoring function $f$ is said to be \textit{monotone} if, for any tuples $s,t$ over $R$,
\begin{equation}
  t[A_i] \leq s[A_i] \Rightarrow f(t) \leq f(s), \forall i=1, \dots, d.
\end{equation}
In \cite{ciaccia_flexible_2020, ciaccia_reconciling_2017}, the authors introduce the $\mathcal{F}$-dominance concept: let $\mathcal{F}$ be a family of scoring functions, then tuple $t$ $\mathcal{F}$-dominates another tuple $s \neq t$, namely $t \prec_{\mathcal{F}} s$, if $$f(t) \leq f(s), \forall f \in \mathcal{F}.$$ The family of all the monotone scoring functions is denoted as $\texttt{MF}$.
\begin{example}\label{example:one}
  Consider three tuples, $t_1 = \langle 1,0 \rangle, t_2 = \langle 0.5, 0.5 \rangle, t_3 = \langle 0,1\rangle$, with dataset dimension $d=2$, and consider the monotone scoring functions $f_1(x,y) = x + y$, $f_2(x,y) = x$. Then, setting $\mathcal{F} = \left\{ f_1, f_2 \right\}$, we have
  \begin{inparaenum}[\itshape a\upshape)]
  \item $t_2 \prec_{\mathcal{F}} t_1$, since $f_1(t_2) = f_1(t_1) = 1$ and $f_2(t_2) = 0.5 < f_2(t_1) = 1$,
  \item $t_3 \prec_{\mathcal{F}} t_1$, since $f_1(t_3) = f_1(t_1) = 1$ and $f_2(t_3) = 0 < f_2(t_1) = 1$, and
  \item $t_3 \prec_{\mathcal{F}} t_2$, since $f_1(t_3) = f_1(t_2) = 1$ and $f_2(t_3) = 0 < f_2(t_2) = 0.5$.
  \end{inparaenum}\\
\end{example}\\

\begin{figure}[t]
\centering
\begin{tikzpicture}
  \begin{axis}[
      axis lines = left,
    xmin = 0, xmax = 1,
    ymin = 0, ymax = 1,
    xlabel = \(A_1\),
    ylabel = {\(A_2\)},
    ytick distance = 0.10,
    xtick distance = 0.10,
    ]

    \addplot [
    dashed,
    domain=0.10:1, 
    samples=200,
    name path=f,
    color=black]
{0.85 - x};

\addplot [
  dashed,
    color=black]
coordinates{(0.10,0.75) (0.10, 1)};

\addplot [
    dashed,
    domain=0.30:1, 
    samples=100,
    name path=h,
    color=black]
{0.80 - x};

\addplot [
  dashed,
    color=black]
coordinates{(0.30,0.50) (0.30, 1)};

\addplot [
    dashed,
    domain=0.50:1, 
    samples=100,
    name path=k,
    color=black]
{0.65 - x};

\addplot [
  dashed,
    color=black]
coordinates{(0.50,0.15) (0.50, 1)};

\addplot [
    dashed,
    domain=0.85:1, 
    samples=100,
    name path=z,
    color=black]
         {0.95 - x};

\addplot [
  dashed,
    color=black]
coordinates{(0.85,0.10) (0.85, 1)};

\addplot [
    dashed,
    domain=0.75:1, 
    samples=100,
    color=black]
coordinates{(0.1, 1) (1,1)};

\addplot [
    dashed,
    domain=0.75:1, 
    samples=100,
    color=black]
coordinates{(1, 0) (1,1)};

\addplot [
    dashed,
    domain=0.75:1, 
    samples=100,
    color=gray]
coordinates{(0.1, 0.75) (0.5,0.15)};

\addplot [
    name path=g,
    color=white]
         {1};

\addplot [gray!20, opacity=0.8] fill between[of=f and g, soft clip={domain=0.10:1}];

\addplot [gray!20, opacity=0.8] fill between[of=h and g, soft clip={domain=0.30:1}];

\addplot [gray!20, opacity=0.8] fill between[of=k and g, soft clip={domain=0.50:1}];

\addplot [gray!20, opacity=0.8] fill between[of=z and g, soft clip={domain=0.50:1}];

\addplot[
    color=black,
    only marks,
    mark=*,
    visualization depends on=\thisrow{alignment} \as \alignment,
    nodes near coords,
    point meta=explicit symbolic,
    every node near coord/.style={anchor=\alignment}
    ]
    table [
      meta index=2
    ] {
      x       y       label    alignment
      0.10    0.75    $t_1$    -40
      0.30    0.50    $t_2$    -40
      0.50    0.15    $t_3$    -10
      0.85    0.10    $t_4$    -160
        };
\end{axis}
\end{tikzpicture}
\caption{Tuples in $[0,1]^2$ and $\mathcal{F}$-dominance regions in gray, where the scoring function is $f(t) = w_1t[A_1] + w_2t[A_2]$ subject to the constraints $\mathcal{C} = \{ w_1 \geq w_2\}$.}
\label{fig:tuples}
\end{figure}
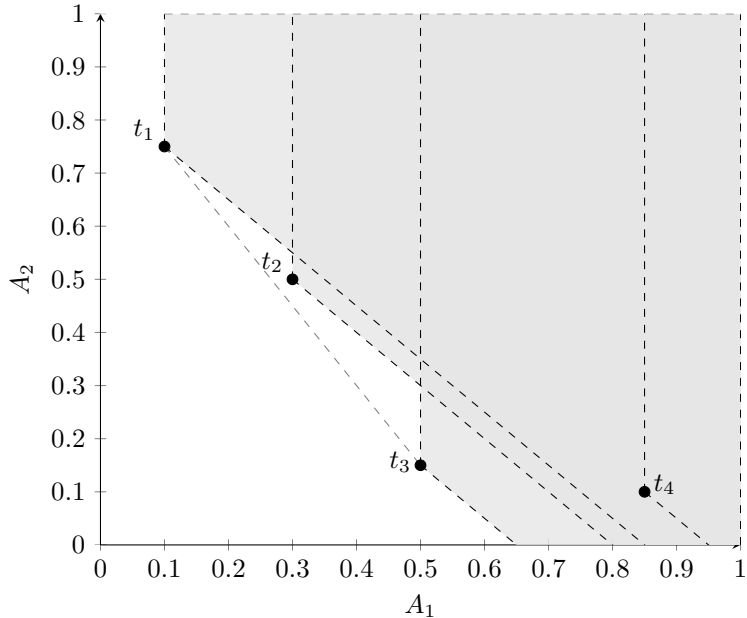

The most interesting case to consider is when the family of linear scoring functions $\mathcal{F}$ is an infinite collection of functions subject to a certain set of linear constraints on weights $\mathcal{C}$. The next example shows how the $\mathcal{F}$-dominance relationships behave in such a scenario.

\begin{example}\label{ex:two}
  Let us consider $\mathcal{F}$ as the set of all scoring function of the form $f(t) = w_1t[A_1]+w_2t[A_2]$, subject to the constraints $\mathcal{C} = \left\{ w_1 \geq w_2 \right\}$. Now, consider the tuples in Figure \ref{fig:tuples}, $t_1 = \langle 0.1, 0.75 \rangle, t_2 = \langle 0.3, 0.5 \rangle, t_3 = \langle 0.5, 0.15 \rangle, t_4  = \langle 0.85, 0.1 \rangle$. We can observe that $f(t_1) = w_1 \cdot 0.1 + w_2 \cdot 0.75 \leq w_1 \cdot 0.3 + w_2 \cdot 0.5 = f(t_2)$ reduces to $w_1 \cdot 0.2 \geq w_2 \cdot 0.25$, which is not true in general. This implies that $t_1 \nprec_{\mathcal{F}} t_2$. With the same argument we can infer that $t_3 \prec_{\mathcal{F}} t_4$, since the inequality reduces to $0.35 \cdot w_1 \geq 0.05 \cdot w_2$, which is true since $w_1 \geq w_2$.
\end{example}

To introduce the F-skyline operators \footnote{In \cite{ciaccia_reconciling_2017} they have a slightly different notation and they are called R-skyline operators.}, two sets need to be defined, as the authors did in \cite{ciaccia_flexible_2020}. The first one is the set of non-dominated tuples wrt. the scoring functions in $\mathcal{F}$, called $\operatorname{ND}$, and the last one is the set of potentially optimal tuples wrt. the scoring functions in $\mathcal{F}$, called $\operatorname{PO}$.\\
In particular, let $\mathcal{F} \subseteq \texttt{MF}$ be a set of monotone scoring functions, then the \textit{non-dominated flexible skyline} of a generic instance $r$ over a schema with attributes $A_1, \dots, A_d$ with respect to $\mathcal{F}$ is the set of tuples
\begin{equation}
  \operatorname{ND}(r; \mathcal{F}) = \left\{ t \in r | \nexists s \in r : s \prec_{\mathcal{F}} t \right\}.
\end{equation}
The \textit{potentially optimal flexible skyline} of $r$ with respect to $\mathcal{F}$ is the set of tuples:
\begin{equation}
  \operatorname{PO}(r; \mathcal{F}) = \left\{ t \in r | \exists f \in \mathcal{F} : \forall s \in r, s \neq t \Rightarrow f(t) < f(s) \right\}.
\end{equation}

To operatively check the $\mathcal{F}$-dominance relationship between tuples, the concept of $\mathcal{F}$-\textit{dominance region} is crucial. In particular, the $\mathcal{F}$-dominance region of a tuple $t$ under a set $\mathcal{F}$ of monotone scoring functions, namely $\operatorname{DR}(t; \mathcal{F})$, is the set of all points $\mathcal{F}$-dominated by $t$:
\begin{equation}
  \operatorname{DR}(t; \mathcal{F}) = \left\{ s \in [0,1]^d : t \prec_{\mathcal{F}} s \right\}.
\end{equation}

It has been proved \cite{ciaccia_flexible_2020} that $\operatorname{ND}$ and $\operatorname{PO}$ are monotone operators wrt. the set of scoring functions, and it has been also proved that for any two sets of monotone scoring functions $\mathcal{F}_1 \subseteq \mathcal{F}_2$ and tuple $t$ over $R$, $\operatorname{DR}(t; \mathcal{F}_1) \supseteq \operatorname{DR}(t; \mathcal{F}_2)$. Intuitively, if the scoring functions in $\mathcal{F}_1$ have stricter constraints they dominate more tuples than the scoring functions in $\mathcal{F}_2$, yielding a broader dominance region.

\subsection{Computing $\operatorname{ND}$ and $\operatorname{PO}$}
The very first problem to solve to compute $\operatorname{ND}$ and $\operatorname{PO}$ is deciding how to check the $\mathcal{F}$-dominance relationships between tuples. The first approach proposed in \cite{ciaccia_reconciling_2017, ciaccia_flexible_2020} is a linear programming (LP) problem in form of minimization. However, computing $\operatorname{ND}$ in that way implies solving a different LP problem for each $\mathcal{F}$-dominance test, entailing a very time-consuming task.\\
Relying on the dominance region concept, instead, allows to compute the set of non-dominated tuples by
\begin{inparaenum}[\itshape a\upshape)]
\item computing the $\mathcal{F}$-dominance regions of tuples and
\item checking whether tuples belong to at least one of such regions. If so, they can be discarded.
\end{inparaenum}\\
The fundamental observation here is that the $\mathcal{F}$-dominance region of a tuple $t$ needs to be computed just once, independently of how many $\mathcal{F}$-dominance tests need to be performed involving tuple $t$. The computation of the dominance region becomes a question of dealing with a convex polytope contained in the standard $(d-1)$-simplex. The most expensive component in obtaining such regions is the vertex enumeration of a polytope, paid once for all tuples.\\
To computation of $\operatorname{PO}(r; \mathcal{F})$, instead, can be executed starting from $\operatorname{ND}(r; \mathcal{F})$, since the following relation always holds:
\begin{equation}
  \operatorname{PO}(r; \mathcal{F}) \subseteq \operatorname{ND}(r; \mathcal{F}) \subseteq \operatorname{SKY}(r).
  \end{equation}
The potential optimality test can be performed by retaining only those tuples that are not $\mathcal{F}$-dominated by any possible convex combination of the remaining ones participating in $\operatorname{ND}$. This is done by checking that a certain linear system is unsatisfiable.\\

\begin{example}
  Consider again the tuples in Figure \ref{fig:tuples} and the instance $r = \left\{ t_1, t_2, t_3, t_4 \right\}$. Let us examine the same family of linear scoring functions as in Example \ref{ex:two}, having the form $f(t) = w_1t[A_1] + w_2t[A_2]$ and being subject to same set of constraints $\mathcal{C} = \left\{ w_1 \geq w_2 \right\}$.
  We can observe immediately that $\operatorname{SKY}(r) = \left\{t_1, t_2, t_3, t_4\right\}$, since there is no tuple in $r$ being dominated by any other tuple in $r$.
  We have that $\operatorname{ND}(r, \mathcal{F}) = \left\{ t_1, t_2, t_3 \right\}$, since tuples $t_1, t_2, t_3$ do not lie in the region $\bigcup_{t \in r} \operatorname{DR}(t;\mathcal{F})$ depicted in gray in Figure \ref{fig:tuples}.
  Finally, since $t_2$ is dominated by a ``virtual'' tuple obtained as a convex combination of $t_1$ and $t_3$, we have that $\operatorname{PO}(r; \mathcal{F}) = \left\{ t_1, t_3 \right\}$.
  We can also observe that
  $$
  \operatorname{PO}(r; \mathcal{F}) = \left\{ t_1, t_3 \right\} \subset \operatorname{ND}(r; \mathcal{F}) = \left\{ t_1, t_2, t_3 \right\} \subset \operatorname{SKY}(r) = r.
  $$
  
\end{example}

Authors propose several algorithms in \cite{ciaccia_flexible_2020} to compute both $\operatorname{PO}$ and $\operatorname{ND}$, with different approaches. Algorithms to compute $\operatorname{ND}$ differ based on three different factors:
\begin{itemize}
\item \textbf{Phases}: choosing whether the computation of $\operatorname{ND}(r; \mathcal{F})$ should be performed after computing the skyline or whether it should be performed directly (respectively, in two phases or one phase);
\item \textbf{Sorting}: choosing whether to sort the dataset before computing $\operatorname{ND}(r; \mathcal{F})$ to produce a topological sort between tuples based on the $\mathcal{F}$-dominance relation (if a tuple $t$ precedes another tuple $s$, then $s$ can't $\mathcal{F}$-dominate $t$);
\item \textbf{$\mathcal{F}$-dominance}: choosing which strategy to apply to compute the $\mathcal{F}$-dominance tests, i.e., whether to solve the LP problem or exploit the dominance region concept and compute the vertex enumeration of a polytope.
\end{itemize}
Different approaches are provided also for the computation of $\operatorname{PO}$:
\begin{itemize}
\item \textbf{Phases}: choosing whether the computation of $\operatorname{PO}(r; \mathcal{F})$ should be performed after computing $\operatorname{ND}(r; \mathcal{F})$ and then filtering out non-$\operatorname{PO}(r; \mathcal{F})$ tuples, or whether it should be performed discarding non-$\operatorname{PO}(r; \mathcal{F})$ tuples directly from the original instance $r$;
\item \textbf{PO test}: choosing whether to compute $\operatorname{PO}(r; \mathcal{F})$ using the primal or the dual $\operatorname{PO}$ test;
\item \textbf{Incrementality}: choosing whether to test the potential optimality of a tuple using an incremental approach (i.e., solving the LP problems of increasing size) or whether to test directly the full instance.
\end{itemize}
Experiments show that, for computing $\operatorname{ND}$, 
\begin{inparaenum}[\itshape a\upshape)]
\item sorting the dataset beforehand always entails better performances,
\item computing the set of non dominated tuples starting from the skyline (i.e., using the two-phases approach) is not beneficial when dealing with large datasets;
\item the vertex enumeration approach is way better than the LP approach in terms of performances.
\end{inparaenum}\\
Experiments show that, to obtain $\operatorname{PO}$, instead,
\begin{inparaenum}[\itshape a\upshape)]
\item computing $\operatorname{ND}$ beforehand and then filtering non-$\operatorname{PO}$ tuples yields to better performances;
\item using the dual form of the LP to check perfect optimality is the better alternative;
\item in challenging scenarios, using the incremental approach is preferable.
\end{inparaenum}

\section{Regret minimizing and k-regret minimizing sets}\label{sec:regret_minimizing}
\subsection{First definitions and properties}\label{subsec:regret_definitions}
One of the well-known drawbacks of top-$k$ queries is the assumption that a utility function used to reduce the original multi-objective problem in a single-objective one always exists. The skyline operator, instead, returns, given only a set of criteria, everything in which the final user may be interested, without limiting the output cardinality.\\
The \textit{k-representative regret minimization query} (k-regret) aims to provide a better way to return a set of representative tuples potentially interesting for the majority of the users involved. In particular, this operator exploits features from both top-$k$ and skyline queries: it returns a set of $k$ tuples without asking the user for a utility function. To reach this goal, in \cite{nanongkai_regret-minimizing_2010}, Nanongkai et al. define, informally, that a user is $x\%$ happy in seeing the $k$ tuples in output if the utility she gets from them is at least $x\%$ of the utility she would obtain seeing the best tuple in the whole database.\\
In \cite{nanongkai_regret-minimizing_2010, chester_computing_2014}, the user-defined scoring function is called \textit{utility function}, but it is similar to the definition given in section \ref{sec:restricted_skylines}. The main difference in defining the regret minimization concept is that, from a user perspective, the higher the attribute values the better (the scoring function is in the utility form).\\
Let $f:\mathbb{R}^+ \rightarrow \mathbb{R}$ be a scoring function of a user, and let $S \subseteq D$ be a set of points. The \textit{gain} of the user is defined as the maximum score derived from $S$, namely
\begin{equation}
  \operatorname{gain}(S, f) = \max_{p \in S} f(p).
\end{equation}

\begin{example}
  Let us now consider again the tuples in Example \ref{example:one}, $t_1 = \langle 1,0 \rangle, t_2 = \langle 0.5, 0.5 \rangle, t_3 = \langle 0,1\rangle$, with a dataset $D$ having dimension $d=2$. Suppose that $S=\{t_1\}$, and suppose that a user has the scoring function $f(x,y) = y$. In this case, since higher attribute values are better than lower ones, $\operatorname{gain}(D,f)=1$, while $\operatorname{gain}(S,f) = 0$.
\end{example}\\

The other two crucial definitions given in \cite{nanongkai_regret-minimizing_2010}, are the definitions of \textit{regret} and \textit{regret ratio}.\\
Let us consider a set of points $S \subseteq D$ and a scoring function $f:\mathbb{R}^+ \rightarrow \mathbb{R}$. The \textit{regret} is defined as
\begin{equation}
  r_D(S, f) = \operatorname{gain}(D,f)-\operatorname{gain}(S,f).
\end{equation}
The regret measures the distance (in score) between the best tuple derived from the whole dataset and the best tuple in the subset $S$.\\
The \textit{regret ratio}, instead, is the measure of the happiness of a given user, namely
\begin{equation}
  rr_D(S,f) = \frac{r_D(S,f)}{\operatorname{gain}(D,f)} = 1 - \frac{\operatorname{gain}(S,f)}{\operatorname{gain}(D,f)}.
\end{equation}

From the definition itself, we can derive that $0 \leq rr_D(S,f) \leq 1$, and the more the regret ratio is closer to $0$ the more the user will be happy.\\
In real-world scenarios users may not know their own scoring functions. In \cite{nanongkai_regret-minimizing_2010}, authors assume that every user has a scoring function of choice, thus considering $F$ as the set of all possible scoring functions. The \textit{maximum regret ratio} is the worst possible regret for any user having her scoring function in $F$:
\begin{equation}
  rr_D(S,F) = \sup_{f \in F} rr_D(S,f) = \sup_{f \in F} \frac{\operatorname{gain}(D,f) - \operatorname{gain}(S,f)}{\operatorname{gain}(D,f)} = \sup_{f \in F} \frac{\max_{p \in D}f(p)-\max_{p \in S}f(p)}{\max_{p \in D} f(p)}.
\end{equation}
As in section \ref{sec:restricted_skylines}, we focus on the family of linear scoring functions, denoted as $\mathcal{L}$.

\begin{example}\label{ex:regret}
  Consider the hotel database in Table \ref{tab:hotel_database} and the scores in Table \ref{tab:hotel_scores}. We have that $\operatorname{gain}(D;f_{(0.6,0.4)}) = 89.0$ (achieved by $t_1$), while $\operatorname{gain}(\{t_3, t_5\}, f_{(0.6,0.4)}) = 82.4$ (achieved by $t_3$). Furthermore, if $S = \left\{ t_3, t_5 \right\}$ we have $r_D(S,f_{(0.6,0.4)}) = 89.0-82.4 = 6.6$ and $rr_D(S,f) = \frac{6.6}{89.0} = 0.07$.
  To find the maximum regret ratio, let us observe that $rr_D(S, f_{(0.6,0.4)}) = 0.07, rr_D(S, f_{(0.8,0.2)}) = 0.09, rr_D(S, f_{(0.4,0.6)}) = 0.05, rr_D(S, f_{(0.2,0.8)}) = 0.08$, thus $rr_D(S,F) = 0.09$, with $F=\left\{ f_{(0.6,0.2)}, f_{(0.8, 0.2)}, f_{(0.4, 0.6)}, f_{(0.2, 0.8)} \right\}$.
\end{example}

\begin{table}[t]
  \caption{Hotels database}
  \centering
  \begin{tabular}{|l|l|l|}
    \hline
    Name & Position rating & Value-for-money rating\\ 
    \hline\hline
    ($t_1$) proArte & 95 & 80 \\
    ($t_2$) easyHotel & 93 & 82\\
    ($t_3$) Atrium & 84 & 80\\
    ($t_4$) ibis & 83 & 89\\
    ($t_5$) Pension Tempel & 82 & 78\\
    \hline
  \end{tabular}
  \label{tab:hotel_database}
\end{table}

\begin{table}[t]
  \caption{Hotel scores}
  \centering
  \begin{tabular}{|l|l|l|l|l|}
    \hline
    Hotel & $f_{(0.6,0.4)}$ & $f_{(0.8, 0.2)}$ & $f_{(0.4, 0.6)}$ & $f_{(0.2,0.8)}$\\
    \hline\hline
    $t_1$ & 89.0 & 92.0 & 86.0 & 83.0\\
    $t_2$ & 88.6 & 90.8 & 86.4 & 84.2\\
    $t_3$ & 82.4 & 83.2 & 81.6 & 80.8\\
    $t_4$ & 85.4 & 84.2 & 86.6 & 87.8\\
    $t_5$ & 80.4 & 81.2 & 79.6 & 78.8\\
    \hline
  \end{tabular}
  \label{tab:hotel_scores}
\end{table}

\subsection{Finding a set that satisfies the ``happiness constraint''}
It has been proved \cite{nanongkai_regret-minimizing_2010} that the maximum regret ratio $rr_D(S, \mathcal{L})$ has \textit{scale invariance} and \textit{stability} properties.\\
A function is said to be scale invariant if rescaling the attribute values (multiplying them by a constant) doesn't change the final solution. In particular, if we rescale the attributes of the dataset $D$ to be analyzed, for instance converting HP to watts, the outcome of the function will remain unvaried.
The stability property, instead, means that a function doesn't change its result if adding non-optimal points for any scoring function in the domain being considered. In other words, the function is insensitive to the addition or deletion of so-called junk points. It has been proved that the regret ratio $rr_D(S, \mathcal{L})$ is a stable function.\\
The stability property is really desirable in several real-world scenarios: a car dealer, for instance, can't manipulate the maximum regret ratio given a certain database by adding non-interesting cars inside of it, since the operator $rr_D(S, \mathcal{L})$ will output the very same result even if junk objects are added into this database.\\
In \cite{nanongkai_regret-minimizing_2010}, two theorems to prove the existence of a lower and upper bound for the maximum regret ratio are provided. Along with these theorems, two algorithms called $\operatorname{CUBE}$ and $\operatorname{GREEDY}$ are described, to find a subset $S$ satisfying, respectively, the upper and lower bound. Nanongkai et al. conjectured that finding an optimal solution of the maximum regret ratio problem is NP-hard. In subsection \ref{subsec:extending_regret}, we describe the extension of the regret minimization problem through the definition of a new operator, the k-regret ratio.

\subsection{Extending the regret minimizing set through the k-regret minimization concept}\label{subsec:extending_regret}
In \cite{chester_computing_2014}, Chester et al. regain the problem introduced by Nanongkai et al. in \cite{nanongkai_regret-minimizing_2010}, extending the regret ratio operator, measuring how far from a $k$-th ``best'' tuple is the best tuple in a subset.\\
Consider a user-specified real-valued vector of attribute weights, called $\mathbf{w}$, and consider a subset $S \subseteq D$, so each tuple $t \in D$ has a score on $\mathbf{w}$, stated as $\operatorname{score}(t, \mathbf{w}) = \sum_{i=1}^d w_i t[A_i] $. The dataset $D$ can be ordered with a descending criterion, producing a list $\left(D^{(1,\mathbf{w})}, \dots, D^{(n,\mathbf{w})}\right)$, so the point $D^{(i,\mathbf{w})}$ is the $i$-th ranked point in $D$ wrt. to the attribute weights defined by the vector $\mathbf{w}$.\\
The $k$gain, the extension of the gain concept introduced in \cite{nanongkai_regret-minimizing_2010}, is the score of the $k$-th ranked point in $S$, namely
\begin{equation}
  k\operatorname{gain}(S, \mathbf{w}) = \operatorname{score}(S^{(k, \mathbf{w})}, \mathbf{w}).
\end{equation}
The $k$-regret ratio, instead, is defined as:
\begin{equation}
  k\operatorname{-regratio}(S, \mathbf{w}) = \frac{\max(0, k\operatorname{gain}(D, \mathbf{w})-1\operatorname{gain}(S, \mathbf{w}))}{k\operatorname{gain}(D, \mathbf{w})}.
\end{equation}
As the regret ratio, the k-regret ratio is such that $0 \leq k\operatorname{-regratio}(S, \mathbf{w}) \leq 1$ always. Note that, setting $k=1$ yields to the definition of regret ratio given in \ref{subsec:regret_definitions}.\\
Considering $\mathcal{L}$ as the set of all possible weight vectors $\mathbf{w} \in [0,1]^d$, the \textit{maximum $k$regret ratio} for a subset $S \subseteq D$ is the maximum possible regret that a user may have looking at the best tuple in the subset $S$ instead of the best $k$ tuples in the whole dataset $D$, namely:
\begin{equation}
  k\operatorname{-regratio(S)}=\sup_{\mathbf{w} \in \mathcal{L}} k\operatorname{-regratio}(S, \mathbf{w}).
\end{equation}
A fixed-size \textit{$k$-regret minimizing set} is a subset $S \subseteq D$ of size $\sigma$ for which the minimum possible maximum $k$-regret ratio is achieved:
\begin{equation}
  R_{\sigma,D} = \operatorname{argmin}_{S \subseteq D, |S| = \sigma} k\operatorname{-regratio}(S).
\end{equation}

\begin{example}
  Consider again the tuples in Example \ref{ex:regret}, shown in Table \ref{tab:hotel_database}, and consider them as normalized in $[0,1]$ and as part of the dataset $D$. The normalized tuples are shown in Table \ref{tab:hotel_normalized} along with their scores for some attribute weights vectors in Table \ref{tab:hotel_weights}. Let us compute the $2$-regret ratio considering the weights vector $ \mathbf{w_3} = \langle 0.4, 0.6 \rangle$. The database can be ordered with a descending criterion wrt. $\mathbf{w_3}$, yielding to $D^{(2,\mathbf{w_3})} = t_2$. Now, considering $S=\{ t_3, t_5 \}$, the $2$-regret ratio will be
  $$
  2\operatorname{-regratio}(S, \mathbf{w_3}) = \frac{\max (0, 0.94-0.89)}{0.94} = 0.05.
  $$
  To find the maximum $2$-regret ratio, instead, let us notice that $D^{(2, \mathbf{w_1})}=t_1, D^{(2, \mathbf{w_2})} = t_2, D^{(2, \mathbf{w_3})}=t_2$. We have
  $$
  2\operatorname{-regratio}(S) = \sup_{\mathbf{w}} 2\operatorname{-regratio}(S, \mathbf{w}) = 0.06,
  $$
  since $2\operatorname{-regratio}(S,\mathbf{w_1}) = 0.05, 2\operatorname{-regratio}(S,\mathbf{w_2}) = 0.06, 2\operatorname{-regratio}(S,\mathbf{w_3}) = 0.05$.
\end{example}

\begin{table}[t]
  \caption{Hotels database normalized in $[0,1]$}
  \centering
  \begin{tabular}{|l|l|l|}
    \hline
    Name & Position rating & Value-for-money rating\\ 
    \hline\hline
    ($t_1$) proArte &  1.00 & 0.89\\
    ($t_2$) easyHotel & 0.98 & 0.92\\
    ($t_3$) Atrium & 0.88 & 0.90\\
    ($t_4$) ibis & 0.87 & 1.00\\
    ($t_5$) Pension Tempel & 0.86 & 0.88\\
    \hline
  \end{tabular}
  \label{tab:hotel_normalized}
\end{table}

\begin{table}[t]
  \caption{Hotels scores} 
  \centering
  \begin{tabular}{|l|l|l|l|}
    \hline
    Hotel & $\mathbf{w_1} = \langle 0.5, 0.5 \rangle $ & $\mathbf{w_2} = \langle 0.6, 0.4 \rangle $ & $\mathbf{w_3} = \langle 0.4, 0.6 \rangle$\\
    \hline\hline
    $t_1$ & 0.94 & 0.96 & 0.93\\
    $t_2$ & 0.95 & 0.95 & 0.94\\
    $t_3$ & 0.89 & 0.89 & 0.89\\
    $t_4$ & 0.93 & 0.92 & 0.95\\
    $t_5$ & 0.87 & 0.86 & 0.87\\
    \hline
  \end{tabular}
  \label{tab:hotel_weights}
\end{table}

In \cite{chester_computing_2014}, Chester et al. proved, reducing the problem to a SET-COVER problem, that finding a set $S$ that achieves the minimum possible maximum $k$-regret ratio is NP-hard. They provide an algorithm for datasets in $d=2$ dimensions based on the convex chain concept to find the $k$-regret minimizing set solution in primal space, and a randomized algorithm for general dimension.

\section{Skyline ordering/ranking}\label{sec:skyline_ordering}
\subsection{Enhancing the skyline flexiblity with size constrained skyline queries}
While the solutions described in previous sections limit the cardinality of potentially interesting tuples, even small skylines could be not completely effective. Consider a classical example of a user booking a hotel room querying a dataset to return the best possible accommodations. The objects belonging to the skyline may refer to hotels fully booked or previously attended by the user with an unsatisfactory experience.\\
In \cite{lu_flexible_2011}, Lu et al. introduce the \textit{size constrained skyline queries} that retrieve $\tilde{k}$ interesting points from a dataset $D$ with dimension $d$, with $\tilde{k}$ that may exceed the skyline cardinality. To obtain such a subset of $\tilde{k}$ tuples, they describe a new approach called \textit{skyline ordering}, a skyline-based partitioning of $D$.\\
A size constrained skyline query, denoted as $Q^{scs}_{\tilde{k}}(D)$, over a $d$-dimensional dataset $D$, is a subset $S \subseteq D$ of $\tilde{k}$ points considered to be \textit{good} in terms of user interest.\\
To make this query effective, authors introduce the skyline ordering. The \textit{skyline order} of a set $D$ consisting of $d$-dimensional data points is a sequence $\mathcal{S} = \langle S_1, \dots, S_n \rangle$ of subsets such that:
\begin{enumerate}
\item $S_1$ is the skyline of $D$;
\item $\forall i, 1 < i \leq n, S_i$ is the skyline of $D \setminus \bigcup_{j=1}^{i-1} S_j$;
\item $\bigcup_{i=1}^n S_i = D$.
\end{enumerate}
Each set $S_i$ is called \textit{skyline order subset} and $n$ is called \textit{skyline order length}, which represents the number of subsets appearing in $\mathcal{S}$.
Informally, each subset $S_i$ is the skyline that would result if the points in the previous sets belonging to the sequence did not exist. We can infer, directly from the definition, that, given the partitions in the skyline order,
\begin{inparaenum}[\itshape a\upshape)]
\item no point belonging to a partition can dominate other points appearing in the same partition or a previous one,
\item every point belonging to a partition, except for the first one, must be dominated by at least one point in the previous partitions.
\end{inparaenum}\\

\begin{figure}[t]
\centering
\begin{tikzpicture}
  \begin{axis}[
      axis lines = left,
    xmin = 0, xmax = 1,
    ymin = 0, ymax = 1,
    xlabel = \(A_1\),
    ylabel = {\(A_2\)},
    ytick distance = 0.10,
    xtick distance = 0.10,
    ]

\addplot [
    dashed,
    domain=0.15:1, 
    samples=100,
    color=black]
coordinates{(0.15, 1) (0.15,0.75)};

\addplot [
    dashed,
    domain=0.15:1, 
    samples=100,
    color=black]
coordinates{(0.15, 0.75) (0.30,0.75)};

\addplot [
    dashed,
    domain=0.15:1, 
    samples=100,
    color=black]
coordinates{(0.30, 0.75) (0.30,0.65)};

\addplot [
    dashed,
    domain=0.15:1, 
    samples=100,
    color=black]
coordinates{(0.30, 0.65) (0.35,0.65)};

\addplot [
    dashed,
    domain=0.15:1, 
    samples=100,
    color=black]
coordinates{(0.35, 0.65) (0.35,0.35)};

\addplot [
    dashed,
    domain=0.15:1, 
    samples=100,
    color=black]
coordinates{(0.35, 0.35) (0.78,0.35)};

\addplot [
    dashed,
    domain=0.15:1, 
    samples=100,
    color=black]
coordinates{(0.78, 0.35) (0.78,0.15)};

\addplot [
    dashed,
    domain=0.15:1, 
    samples=100,
    color=black]
coordinates{(0.78, 0.15) (0.95,0.15)};

\addplot [
    dashed,
    domain=0.15:1, 
    samples=100,
    color=black]
coordinates{(0.43, 1) (0.43,0.70)};

\addplot [
    dashed,
    domain=0.15:1, 
    samples=100,
    color=black]
coordinates{(0.43, 0.70) (0.56,0.70)};

\addplot [
    dashed,
    domain=0.15:1, 
    samples=100,
    color=black]
coordinates{(0.56, 0.70) (0.56,0.57)};

\addplot [
    dashed,
    domain=0.15:1, 
    samples=100,
    color=black]
coordinates{(0.56, 0.57) (0.73,0.57)};

\addplot [
    dashed,
    domain=0.15:1, 
    samples=100,
    color=black]
coordinates{(0.73, 0.57) (0.73,0.40)};

\addplot [
    dashed,
    domain=0.15:1, 
    samples=100,
    color=black]
coordinates{(0.73, 0.40) (0.95,0.40)};

\addplot [
    dashed,
    domain=0.15:1, 
    samples=100,
    color=black]
coordinates{(0.63, 1) (0.63,0.77)};

\addplot [
    dashed,
    domain=0.15:1, 
    samples=100,
    color=black]
coordinates{(0.63, 0.77) (0.90,0.77)};

\addplot [
    dashed,
    domain=0.15:1, 
    samples=100,
    color=black]
coordinates{(0.90, 0.77) (0.90,0.56)};

\addplot [
    dashed,
    domain=0.15:1, 
    samples=100,
    color=black]
coordinates{(0.90, 0.56) (0.95,0.56)};

\node[] 
at (axis cs:0.97,0.15) {$S_1$};

\node[] 
at (axis cs:0.97,0.40) {$S_2$};

\node[] 
	at (axis cs:0.97,0.56) {$S_3$};

\addplot[
    color=black,
    only marks,
    mark=*,
    visualization depends on=\thisrow{alignment} \as \alignment,
    nodes near coords,
    point meta=explicit symbolic,
    every node near coord/.style={anchor=\alignment}
    ]
    table [
      meta index=2
    ] {
      x       y       label    alignment
      0.15    0.75    A        -40
      0.30    0.65    B        -40
      0.35    0.35    C        -40
      0.78    0.15    D        -40

      0.43    0.70    E        -40
      0.56    0.57    F        -40
      0.73    0.40    G        -40

      0.63    0.77    H        -40
      0.90    0.56    I        -40
        };
\end{axis}
\end{tikzpicture}
\caption{Skyline order}
\label{fig:skyline}
\end{figure}
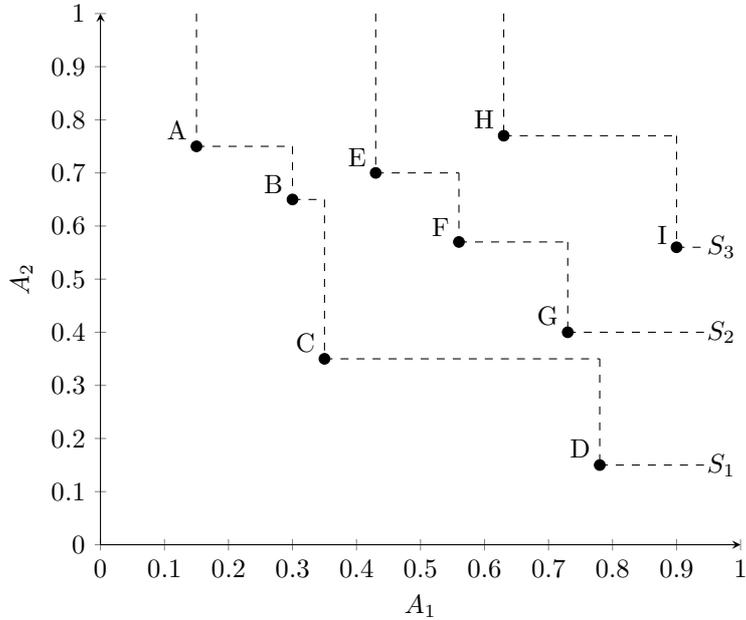

Three properties define how these points belonging to the skyline order are related to each other. Let us consider the skyline order $\mathcal{S} = \langle S_1, \dots, S_n \rangle$ of a $d$-dimensional dataset $D$, the following properties hold:
\begin{enumerate}
  \item $\forall i > 1, \forall s \in S_i, \text{ } \exists t \in S_{i-1}$ such that $t \prec s$;
  \item $\forall i > 1, \forall s \in S_i, \forall j < i, \text{ } \exists t \in S_j$ such that $t \prec s$;
  \item $\forall s \in S_i, \forall j > i, \text{ } \nexists t \in S_j $ such that $t \prec s$.
\end{enumerate}

\begin{example}
  Consider the points in Figure \ref{fig:skyline}. The set of points $S_1 = \left\{ A, B, C, D \right\}$ is the skyline of all tuples, while $S_2 = \left\{E, F, G \right\}$ is the skyline for subset $\left\{ E, F, G, H, I \right\}$. Finally, $S_3 = \left\{ H, I \right\}$ is the skyline of subset $\left\{ I, H \right\}$. The three skyline order subsets are shown in Figure \ref{fig:skyline}.
\end{example}

Considering the example regarding the room accommodations described before, applying the skyline-based partitioning can be extremely effective to satisfy most of the user's requirements: if the standard skyline returns a set of non-satisfactory tuples, for any of the possible reasons described before, a broader selection can be provided, relying on the skyline with order $2$.\\
In \cite{lu_flexible_2011}, the authors provide an algorithm for the skyline order computation, which checks for each point belonging to the dataset $D$ its affinity with each skyline order subset, thus creating a new subset if no already existent subset is suitable for the point taken into account. This algorithm can be improved by presorting the dataset $D$ in non-descending order wrt. each point's attribute values, and by carrying out a binary search instead of looping over all the possible skyline order subsets to find a suitable one for the current tuple to be analyzed. Finally, an algorithm is provided to process the size constrained skyline query itself, in two variants, one with the precomputed skyline order and one computing it directly.

\subsection{Ranking the skyline points through the skyline graph}
Another similar approach has been proposed in \cite{vlachou_ranking_2010}, through a framework called SKYRANK, which aims to rank the skyline points to obtain a small subset of interesting ones. This framework works without the need to have a user-defined scoring function, even if an extension working with available user preferences is provided.\\
The skyline set is built using the dominance relationship, which is applied considering the set of all attributes defined over the dataset $D$. The same relationship can be considered with respect to different subsets of attributes \cite{10.5555/1083592.1083624}, called \textit{subspaces}, and such a relationship
is used to define the ranking approach. The subspace dominance relationships between points in $D$ are mapped into a weighted directed graph, called the \textit{skyline graph}.\\
Consider a dataset $D$ on a data space $\bar{D}$ in $d$ dimensions, over the attributes $A_1, \dots, A_d$. A tuple $t$ \textit{dominates} another tuple $s$ on subspace $\bar{U} \subset \bar{D}$, $t \prec_{\bar{U}} s$, if
\begin{inparaenum}[\itshape 1\upshape)]
\item on every dimension $d_i \in \bar{U}$, $t[A_i] \leq s[A_i]$,
\item on at least one dimension $d_j \in \bar{U}$, $t[A_j] < s[A_j]$.
\end{inparaenum}\\
The subspace dominance relationship is a ``restriction'' of the usual dominance concept: the inequality checks are performed considering a subset of the whole attributes.
The skyline of a subspace $\bar{U}$ is set of points which are not dominated by any other point lying in subspace $\bar{U}$.

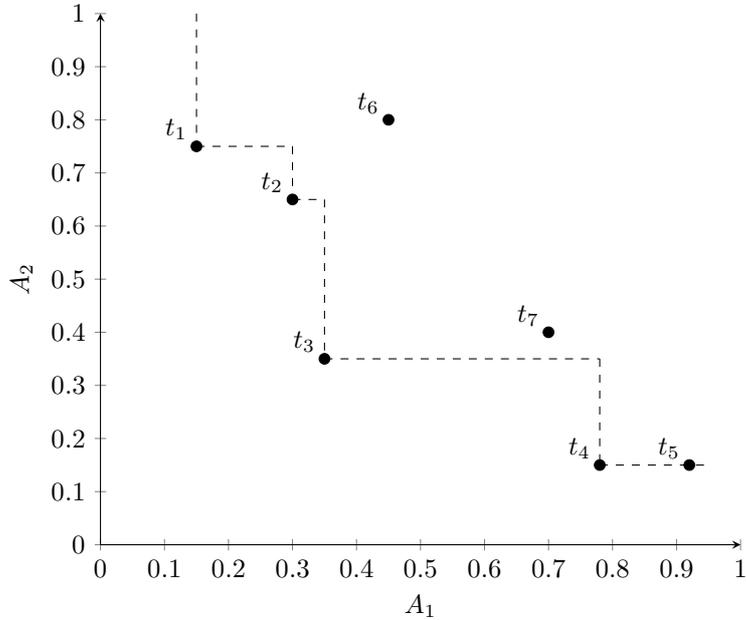
\begin{figure}[t]
\centering
\begin{tikzpicture}
  \begin{axis}[
      axis lines = left,
    xmin = 0, xmax = 1,
    ymin = 0, ymax = 1,
    xlabel = \(A_1\),
    ylabel = {\(A_2\)},
    ytick distance = 0.10,
    xtick distance = 0.10,
    ]

\addplot [
    dashed,
    domain=0.15:1, 
    samples=100,
    color=black]
coordinates{(0.15, 1) (0.15,0.75)};

\addplot [
    dashed,
    domain=0.15:1, 
    samples=100,
    color=black]
coordinates{(0.15, 0.75) (0.30,0.75)};

\addplot [
    dashed,
    domain=0.15:1, 
    samples=100,
    color=black]
coordinates{(0.30, 0.75) (0.30,0.65)};

\addplot [
    dashed,
    domain=0.15:1, 
    samples=100,
    color=black]
coordinates{(0.30, 0.65) (0.35,0.65)};

\addplot [
    dashed,
    domain=0.15:1, 
    samples=100,
    color=black]
coordinates{(0.35, 0.65) (0.35,0.35)};

\addplot [
    dashed,
    domain=0.15:1, 
    samples=100,
    color=black]
coordinates{(0.35, 0.35) (0.78,0.35)};

\addplot [
    dashed,
    domain=0.15:1, 
    samples=100,
    color=black]
coordinates{(0.78, 0.35) (0.78,0.15)};

\addplot [
    dashed,
    domain=0.15:1, 
    samples=100,
    color=black]
coordinates{(0.78, 0.15) (0.95,0.15)};

\addplot[
    color=black,
    only marks,
    mark=*,
    visualization depends on=\thisrow{alignment} \as \alignment,
    nodes near coords,
    point meta=explicit symbolic,
    every node near coord/.style={anchor=\alignment}
    ]
    table [
      meta index=2
    ] {
      x       y       label    alignment
      0.15    0.75    $t_1$    -40
      0.30    0.65    $t_2$    -40
      0.35    0.35    $t_3$    -40
      0.78    0.15    $t_4$    -40

      0.92    0.15    $t_5$    -40
      0.45    0.80    $t_6$    -40
      0.70    0.40    $t_7$    -40
        };
\end{axis}
\end{tikzpicture}
\caption{Subspace dominance relationships}
\label{fig:subspace}
\end{figure}

\begin{example}
  Consider the tuples in Figure \ref{fig:subspace} under the instance $r= \left\{ t_1, t_2, t_3, t_4, t_5, t_6, t_7 \right\}$. We have that $\operatorname{SKY}(r) = \left\{ t_1, t_2, t_3, t_4 \right\}$, while for subspace $\bar{U} = \left\{A_2\right\}$, the subspace skyline is $\operatorname{SKY}_{\bar{U}}(r) = \left\{ t_4, t_5 \right\}$. Note that $t_5$ is a skyline point in subspace $\left\{A_2 \right\}$ but it is dominated in the full space by $t_4$.
\end{example}

Given a generic instance $r$, each subspace skyline tuple $t \in \operatorname{SKY}_{\bar{U}}(r)$ 
\begin{inparaenum}[\itshape a\upshape)]
\item belongs to the skyline on $\bar{D}$, or
\item it is dominated on $\bar{D}$ by another tuple $s \in \operatorname{SKY}_{\bar{U}}(r)$, such that $t[A_i] = s[A_i]$, for each dimension $d_i \in \bar{U}$.
\end{inparaenum}\\
Given a $d$-dimensional dataset $D$, there can be $2^d-1$ possible skylines defined over different subspaces.\\
The skyline graph is built using vertices corresponding to skyline points and adding directed edges $e_{st}$ from $s$ to $t$ if tuple $t$ dominates another tuple $s$ in a subspace of the original dataspace.\\
The \textit{interestingness} of a skyline point $t$ is based on the dominance relationships in different subspaces, and
\begin{inparaenum}[\itshape a\upshape)]
\item a tuple is more interesting if it dominates many other ``important'' skyline tuples in many different subspaces,
\item the interestingness property can be passed from a tuple to all the other tuples that dominate it.
\end{inparaenum}\\
Vlachou et al. propose \cite{vlachou_ranking_2010} a brief and concise formula to measure the interestingness of a skyline tuple $t$:
\begin{equation}
I(t) = (1-\alpha) \sum_{t \in \operatorname{domd}(s)} \frac{1}{|\operatorname{dom}(t)|} I(s) + \alpha \frac{1}{|\operatorname{SKY}|},
\end{equation}
where $\operatorname{domd}(s)$ is the set of tuples that are dominated by $s$ in any subspace, $\operatorname{dom}(t)$ is the set of tuples that dominate $t$ in any subspace, and $\alpha \in [0,1]$ is used to balance between the two terms of the score.\\
Once the algorithm to build the skyline graph is executed, another link-based ranking procedure \cite{brin_anatomy_1998} is applied to assign a score to each vertex. The main idea is that a subspace skyline point $s$ that is dominated by another skyline point $t$ in another subspace, transfers a fraction of its interestingness to the dominating point $t$. Using these link-based ranking techniques, the more some points dominate other points in different subsets the more they will be highly ranked.\\
Considering a real-world scenario, a user finding the best car from the set of the skyline cars according to the manufacturer's website can be interested with high probability also in those cars that dominate the best one in other subspaces, rather than another random car belonging to the original skyline.

\section{Conclusions}\label{sec:conclusions}

\begin{table}[t]
  \caption{Operators and their properties}
  \centering
  \begin{tabular}{|l|l|l|l|}
    \hline
    \textbf{Operator} & \textbf{Personalization} & \textbf{Output-size control} & \textbf{Flexibility}\\ 
    \hline\hline
    Flexible/restricted skylines & \cmark & \xmark & \cmark\\
    Regret minimizing sets & \xmark & \cmark & \cmark\\
    Skyline ordering/ranking & \xmark & \cmark & \cmark\\
    \hline
  \end{tabular}
  \begin{tabular}{|l|l|l|}
    \hline
    \textbf{Operator} & \textbf{Stability} & \textbf{Scale invariance}\\
    \hline\hline
    Flexible/restricted skylines & \cmark & \xmark\\
    Regret minimizing sets & \cmark & \cmark\\
    Skyline ordering/ranking & \cmark & \cmark\\
    \hline
  \end{tabular}
  \label{tab:operator_properties}
\end{table}

Three aspects are commonly considered when evaluating the applicability and effectiveness of multi-objective optimization's frameworks \cite{mouratidis_marrying_2021}. The first one regards whether the operators provide a kind of \textit{personalization} in serving the preferences of the final users. Reporting the same result for every user could be not as effective in a modern real-world scenario, with a large amount of personal information available. Another aspect that is commonly taken into account is the control of the result's cardinality, thus developing \textit{output-size specified} operators. This can be critical for the quality of the decision to be made and also for design reasons (displaying a huge amount of objects is not as easy as displaying a small subset of representative ones). The last criterion commonly considered is \textit{flexibility} in specifying the user's preferences. Asking a user to provide exact preferences (in a top-$k$ scenario, providing weights to apply for each attribute) is a non-trivial operation, and mining those preferences without directly asking has some precision drawbacks.\\
The skyline ordering approaches proposed by Lu et al. \cite{lu_flexible_2011} and by Vlachou et al. \cite{vlachou_ranking_2010} do not consider user preferences, thus providing a subset equal for all users involved. The regret minimization techniques \cite{nanongkai_regret-minimizing_2010, chester_computing_2014} do not support personalization as well. Moreover, getting an optimal solution for the regret minimizing set problem is NP-hard. Personalization is provided, instead, with flexible/restricted skyline techniques \cite{ciaccia_flexible_2020, ciaccia_reconciling_2017} and with the operators described in \cite{mouratidis_marrying_2021, mouratidis_exact_2018}. However, flexible/restricted skylines do not support the output-size control: they reduce the output cardinality but without knowing the exact number of tuples to be returned by the operators. This property is observed, instead, by regret minimizing sets (they return exactly $k$ representative tuples) and by skyline ordering/ranking techniques (with both size constrained skyline queries and with SKYRANK, respectively). All three different approaches provide input flexibility, both avoiding to ask directly to the user for her preferences and considering a broader set of scoring functions (as authors do, for instance, in \cite{ciaccia_flexible_2020, ciaccia_reconciling_2017}).\\
Two further properties that are desirable in plenty of real-world scenarios are scale invariance and stability. We have already discussed their satisfaction with the operators employed in the regret minimization techniques. Flexible/restricted skylines, instead, are stable along with skyline ordering/ranking techniques: when junk objects are added into the database their outcome does not change. However, flexible/restricted skylines are not scale invariant: changing the attributes' values when doing the rescale process can compromise the constraints that have been set on the family of scoring functions taken into account. The skyline ordering/ranking techniques support also the scale invariance property.\\
Table \ref{tab:operator_properties} summarizes the operators' properties described above, to better highlight which of the described approaches is suitable when operating in a certain scenario. Recent approaches \cite{mouratidis_marrying_2021} aim to provide all the three properties described before and efficient algorithms to compute the desired outcome. Some extensions \cite{wang_fully_2021}, instead, aim to enhance some functionalities provided by the described operators and to adapt them in modern real settings.

\bibliographystyle{plain}
\bibliography{main.bib}
\end{document}